\documentclass{article}

\usepackage[english]{babel}

\usepackage[letterpaper,top=2cm,bottom=2cm,left=3cm,right=3cm,marginparwidth=1.75cm]{geometry}

\usepackage{amsmath}
\usepackage{graphicx}
\usepackage[colorlinks=true, allcolors=blue]{hyperref}
\usepackage{xcolor}
\usepackage{float}
\usepackage{authblk}

\newcommand{\mcomment}[2][\void]{\marginpar{\raggedright\tiny\ifx\void#1\else\textbf{#1:\\}\fi#2}}

\definecolor{DodgerUniformBlue}{rgb}{0.0,0.353,0.612}
\newcommand{\define}[1]{\emph{\textcolor{DodgerUniformBlue}{#1}}}

\title{Making Quantum Accessible: A Seven-Category Framework for K-12 Quantum Education}

\author{Rhea Fernandez}
\author{Sarah Hagstrom}
\author{Liesel Malanos}
\author[1,2]{Lachlan McGinness}
\author{Madeline Mitchell}
\author{Saskia Schultz}
\author{Elizabeth Sexton}

\affil[1]{Australian National University}
\affil[2]{Commonwealth Scientific and Industrial Research Organisation}

\date{}

\begin{document}
\maketitle

\begin{abstract}
We conducted a literature review and expert interviews 
to determine the most common methods being used to teach quantum physics and quantum computing concepts to primary and secondary students.
Based on the findings of this review, we provide a framework of seven categories of teaching approaches for teaching mathematically accessible quantum concepts; they are Defamiliarization, Quantum Picturalism, Spin-First Approach, Einstein-First Approach, Many Paths Approach, Historical Development Approach and Game-based Quantum Learning.
We summarise each of these teaching methods and overview their advantages and disadvantages of each method. 
Our framework makes it easy for physics educators to embrace the diverse methods of teaching quantum physics and quantum computing at the primary and secondary level. 

\end{abstract}

\section{Introduction}

Quantum Mechanics is included in many high school curricula including the Australian curriculum \cite{AustralianCurriculumPhysics} and the International Baccalaureate \cite{IB}. Many resources are being developed with little awareness of similar approaches already existing. For example quantum games, there are more than 100 existing online quantum games \cite{Proquest016} and many versions of quantum tic-tac-toe \cite{Proquest010}. 
The purpose of this paper is to make both educators and researchers aware of the many methods that already exist: to enhance their current teaching and prevent `re-inventing the wheel'. 
Our work also serves as a framework to classify and describe different approaches to teaching quantum mechanics, allowing teachers to discover new ways of teaching quantum physics to their students.\\

Our investigation was two-fold: we conducted a literature review of quantum teaching approaches and we conducted online interviews with academics and professionals who teach quantum physics or quantum computing. From this point, we will refer to the people interviewed as quantum experts. \\

The database search took place in March 2025, two sets of search terms were used [`Quantum', `Education'], and [`Quantum', `Teaching']. The databases searched were ERIC, the arXiv, IEEExplore, Proquest, ACM and Springer. This returned a total of 2,077 papers which were collected from each individual database then papers which appeared in multiple databases (duplicates) were removed. Using techniques from \cite{McGinness2024Highlighting} these were both automatically and manually title and abstract screened for relevance, leaving a total of 167 papers. 
To be relevant, a paper should be focused on the teaching of quantum physics or quantum computing and related to pre-tertiary education. During this review we noticed that there is a teaching phenomenon called quantum learning occurring in the United States of America which has nothing to do with quantum physics. Papers on this topic were excluded.
These papers were then read in depth to determine their relevance. We also investigated the references of relevant papers. \\

We also interviewed a total of eleven quantum experts. The experts described their experience teaching quantum physics and quantum computing and also outlined any prominent quantum teaching methods that they were aware of. The experts alerted us to Bob Coecke's Quantum Picturalism approach, Richard Feynman's Many Paths Approach, the Einstein-First approach and named the Spin-First approach. By looking at the papers which did not fit into these categories we created the following categories: Historical Development Approach, Defamiliarsiation and Game-Based Quantum Learning. 

In the following sections we outline each of the categories of quantum teaching. We give a basic introduction to the methods used in the approach and discuss the advantages and limitations of each approach. 

\section{Historical Development Approach}

The Historical Development Approach refers to the method used to teach quantum mechanics in secondary curricula like the Australian Curriculum and International Baccalaureate Diploma. 
In this approach, students are exposed to particular experiments which demonstrate quantum phenomena. 
For example, the photoelectric effect is used to demonstrate the particle nature of light,
the double slit experiment is used to demonstrate the wave nature of light, and
atomic spectra are used to demonstrate discrete energy levels within atoms. 
The Historical Development Approach employs an experimental focus that lends itself to conducting practical experiments that can increase student engagement. 



\begin{figure}[H]
\centering
\includegraphics[width=0.6\linewidth]{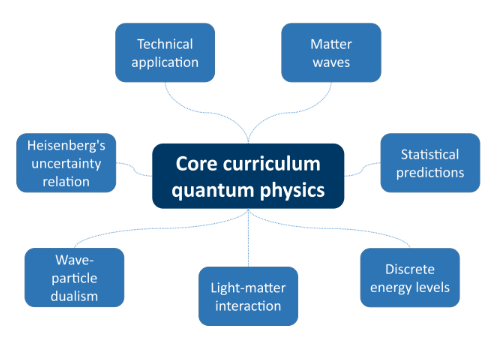}
\caption{An overview of the key topics explored through the Historical Development Approach in secondary schools \cite{Bitzenbaur2021Effect}.}
\end{figure}

\paragraph{}
In the context of high school, the Historical Development Approach offers an effective introduction to quantum mechanics. The framework provides a conceptual explanation of quantum mechanics that does not rely on students having a strong mathematical understanding \cite{Boe2023Secondary}. 
This allows students to understand the development of quantum theories and their implications, even if they find mathematical concepts challenging \cite{Eric105}. 
For example, exploring the counterintuitive results of historical experiments encourages students to discuss how quantum physics may contradict classical physics. This promotes critical engagement with quantum theories \cite{Proquest008}.

\paragraph{}
The Historical Development Approach does not require students to have knowledge of calculus or differential equations, but still allows them to engage with mathematical concepts such as the Heisenberg uncertainty principle, the Rydberg formula, the Planck Law, Wien's Law and photoelectric stopping voltage \cite{Eric092}. 

\paragraph{}
The Historical Development Approach allows students to understand many applications of quantum mechanics including lasers and transistors.
This helps students appreciate the relevance of quantum mechanics for everyday life and hopefully will encourage them to pursue quantum in their later studies. 
However, understanding the applications of quantum physics may also be insufficient to encourage an appreciation for learning about quantum \cite{Eric032}. 
Therefore, this approach normally avoids talking about the Schrodinger equation and the wave-function. 
Quantum circuits including qubits and quantum gates are generally not covered in this approach.

\paragraph{}
Although many different phenomena are covered in this approach, they are not necessarily unified into one consistent picture. This may leave students feeling like they know things about quantum without actually knowing `what quantum mechanics is', as, "Despite formal instruction, learners often develop fragmented ideas or misconceptions about quantum concepts" \cite{Hennig2024Introducing}. 

\paragraph{}
A study by Maria Vetleseter B{\o}e and Susanne Viefers \cite{Boe2023Secondary} found that few secondary students taught by this approach used mathematical concepts to demonstrate their understanding about aspects of quantum physics. This illustrates the challenges students face in connecting mathematical concepts with quantum principles when using this approach.

\section{Defamiliarisation}
Defamiliarisation is a term used to describe media which use comparisons, analogies and metaphors to present the familiar ideas in unfamiliar contexts. Although defamiliarisation normally refers to the arts (like poetry and fiction), the same technique has been widely used to engage the general public in quantum mechanical concepts and theories. 
It effectively breaks down quantum’s counterintuitive nature to those with little knowledge of the subject and is used to convey core concepts such as measurement, superposition, entanglement, and spin. Therefore, this technique is often used in conjunction with other teaching methods due to its simple and versatile nature. 
In contrast to other introductory methods (such as the Spin-First, Einstein-First, or even Quantum Picturalism approach), defamilarisation is usually `non-committal' and can feel more like a one-off exploration rather than a structured curriculum.

\paragraph{} Defamiliarisation texts usually appear in the form of publicly available and widely accessible books, documentaries, and audiovisual presentations.  Well-known examples which take this approach include `Quantum for Babies' by Chris Ferrie \cite{Ferrie2017Quantum}, `A Brief History of Time' by Stephen Hawking \cite{Hawking2016Brief}, `What is Real?' by Adam Becker \cite{Becker2018What}, and The Secrets of Quantum Physics by the BBC \cite{BBC2014Secrets}. \\

Another example is Dominic Walliman’s TedX talk `Quantum Physics for 7 Year Olds' \cite{Walliman2016Quantum}. In this talk Walliman explains key phenomena of quantum mechanics using simple analogies. He likens subatomic particles to the familiar idea of ``bouncy balls'' which forms the basis of his discussion as he introduces particle-wave duality; ``imagine dropping a bouncy ball into a pond of water... the ball would disappear then you’d get these ripples going out over the surface… one of the ripples hits a stick. All of the ripples on the surface disappear and the bouncy ball pops out again.''; quantum tunnelling; he describes the bouncy ball bouncing against a wall multiple times and then spontaneously bouncing through the wall without interacting with it; superposition; for this demonstration, Walliman physically demonstrates spinning around in opposite directions and then tries to imitate what it would look like to spin in both directions at the same time.

\begin{figure}[H]
\centering
\includegraphics[width=0.6\linewidth]{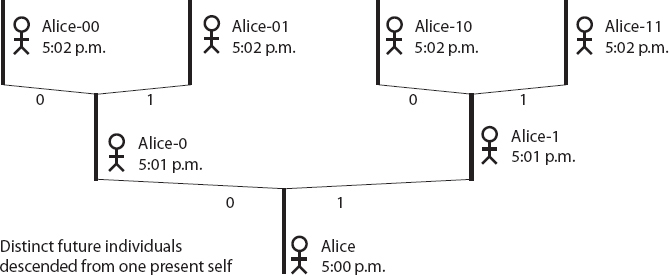}
\caption{Diagram from Something Deeply Hidden \cite{Carroll2021Something} used to explain the Many Worlds interpretation of Quantum Mechanics.}
\label{fig:Alice}
\end{figure}

\paragraph{} Similarly in Sean Carroll's 2021 book `Something Deeply Hidden' \cite{Carroll2021Something}, he explains the Many Worlds interpretation of quantum mechanics through a simple analogy. 
Carroll uses an analogy to explain that ``in Many-Worlds, the life-span of a person should be thought of as a branching tree, with multiple individuals at any one time, rather than a single trajectory... Rather than talking about the `you at 5:01 p.m.', we need to talk about `the person at 5:01 p.m. who descends from you at 5:00 p.m. and who ended up on the spin up branch of the wave function' and likewise for the spin-down branch.''
Using this analogy and a simple diagram, see Figure \ref{fig:Alice}, Carrol explains the logic of the unfamiliar Many Worlds approach through familiar use of a tree-diagram and different versions of a girl named Alice at different times in the day.

\paragraph{} The advantage of the Defamiliarisation approach it it's lack of mathematical pre-requisites making it a highly accessible method of learning quantum concepts for young people or those without a background in higher mathematics and physics. Furthermore, there are many easily accessible materials in a wide range of modes and media that take this approach. This includes books and documentaries written by experts, guided activities and games, and thought experiments.

However, the Defamiliarisation approach can only communicate the most elementary aspects of quantum mechanics as it tends away from mathematical formalism. In some cases, this deviation may create inaccuracies and misconceptions that are counterproductive to the learning process and must be avoided. Furthermore, there are little to no practical takeaways which limit the approach to a short-term introduction to quantum mechanics.

\section{Quantum Picturalism}
Quantum Picturalism is a teaching method developed by Physicist and Professor Bob Coeke in 2009, aiming to teach foundational quantum computing concepts in a visual and accessible way. Some of these key concepts include: wires (qubits) and boxes (gates), tests, teleportation and entanglement, spiders (measurement), phases, sure boxes, states, superposition, and quantum uncertainty. 

\paragraph{} Quantum Picturalism is based on a form of graphical mathematics called ZX-calculus which was developed in 2007 by Bob Coeke and Ross Duncan \cite{Coecke2013Tutorial}. The method uses a non-mathematical approach and therefore has a greater focus on diagrams, analogies and greater conceptual explanations, concentrating on the observation and understanding of concepts at a foundational level before they are fully fleshed-out and conventionally named. 

\begin{figure}
  \centering
\includegraphics[width=0.8\linewidth]{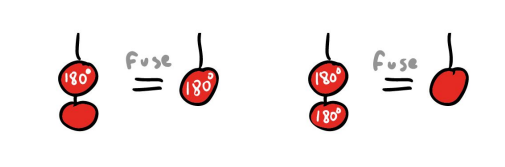}
  \caption{Illustration of phase arithmetic using ZX-calculus spiders, adapted from `Quantum in Pictures' \cite{Coecke2023Quantum}. The image depicts the `fuse' rule, where two connected spiders combine. The upper $180^\circ$ spider acts as a toggle operator on the lower spider's phase: it flips a $0^\circ$ phase to $180^\circ$, and a $180^\circ$ phase to $360^\circ$ (equivalent to $0^\circ$).}
  \label{fig:picturalism} 
\end{figure}

As for the approach itself, Coeke uses diagrams depicting wires and boxes (qubits and gates) and demonstrates the physical manipulation of these by cutting wires and rotating boxes, demonstrating real concepts and illustrating symmetries. By using a visual aid alongside a relevant explanation, as in Figure \ref{fig:picturalism}, students can better understand and learn to build quantum circuits and understand broader concepts. 
The visual nature of the diagram in Figure \ref{fig:picturalism}, for example, is helpful in determining what exactly happens when a NOT Gate is used on a qubit. Rather than seeing a simple number change, a visual alteration occurs. This also means that in more complicated or extensive circuits, the process by which the solution is derived can be visually mapped out and traced (fusion, copying and gates applied to spiders) rather than attempting to understand lines of code and lists of numbers which have led to a solution. 

\paragraph{} The main texts which utilise Quantum Picturalism are: `Quantum in Pictures' \cite{Coecke2023Quantum} which uses predominantly diagrams and simple explanations and includes specialised terms and analogies for better understanding (widely accessible to many demographics); 
and `Picturing Quantum Processes' \cite{Coecke2017Picturing} which is similar to `Quantum in Pictures' but targeted at undergraduate physics students. 
Coeke defines Quantum Picturalism as ``Learning in the Language of Diagrams'' \cite{Coecke2017Picturing}. Because of the diagrammatic nature of the technique, the method does not require prerequisite mathematical knowledge in order for quantum concepts to be learned. Quantum Picturalism is mainly targeted towards high-school and undergraduate students, but it can be adapted for many audiences.

\paragraph{} The main advantage of Quantum Picturalism lies in its capacity to produce a strong baseline understanding of quantum computing concepts in a conceptual way, making it more accessible to a general audience. This can be used as the foundation for learning more complex quantum concepts. 
The main disadvantage of Quantum Picturalism as an approach to teaching is that it only focuses on quantum computing and quantum circuits and does not necessarily cover wider quantum phenomenon. 

\section{Spin First Approach}
\label{sec:SpinFirst}
The Spins First Approach is commonly used to introduce quantum computing and quantum information science at a high school level \cite{Sadaghiani2015Spin,Quantum101}.
The approach is characterised by its focus on conceptual understanding of qubits as the basis of quantum computers. It first discusses qubits alongside gates and the Bloch Sphere, before expanding into more complex topics such as algorithms and qubit manifestations \cite{ERIC142}. Quantum computers are compared to their counterparts in classical computing. 

\paragraph{} The Spins First Approach allows students to deepen their understanding of quantum concepts through code as it enables them to run simulated quantum programs. This approach uses quantum computers as a vehicle to teach four key concepts; superposition, measurement, entanglement and interference \cite{Merzel2024Mathematical}. By exploring these Quantum Computing concepts, the approach encourages students to speculate about further Quantum Phenomena. This ultimately introduces  students to more complex Quantum Mechanics.

\begin{figure}[ht]
\centering
\includegraphics[width=0.7\linewidth]{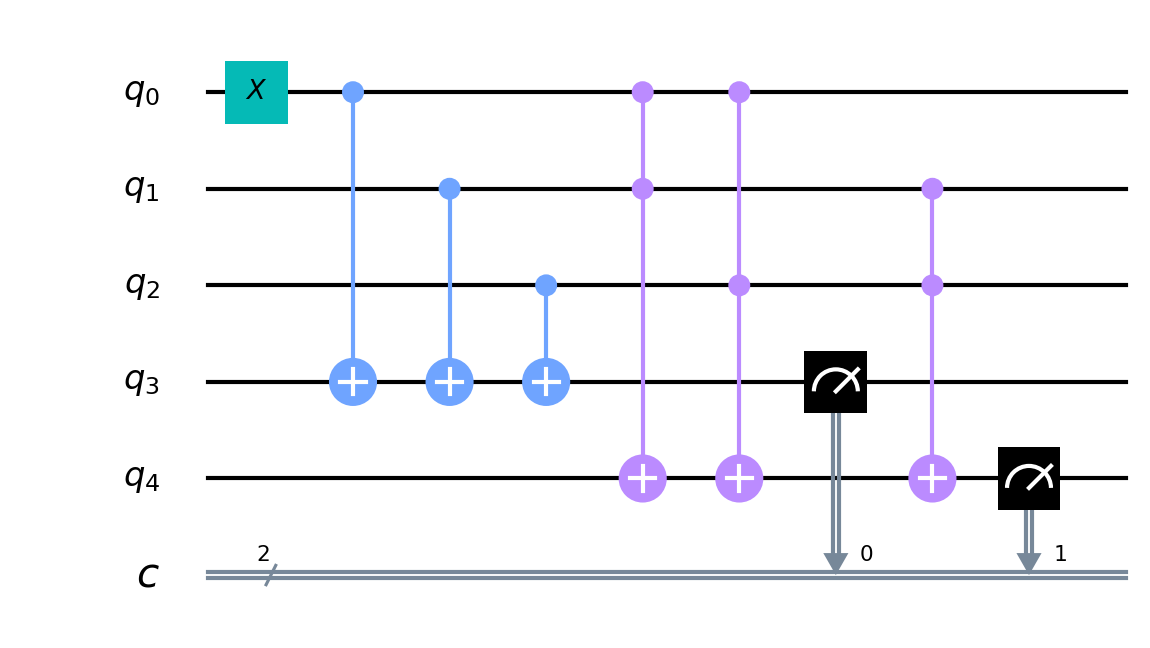}
\caption{\label{fig:QuantumCircuit} This is a diagram of a quantum circuit, represented through the gate based model \cite{Coggins2019Performing}. On the y-axis there are qubits labeled with names such as $q_0$, with a control qubit at the bottom. The x-axis can be thought of as time where the different gates are applied to qubits in the indicated order.}
\end{figure}

\paragraph{} The Spins First Approach teaches the quantum mechanics required for quantum information science without the need for university level mathematics. This approach relies on a student’s pre-established knowledge of geometry, while teaching the concepts of vectors, complex numbers and tensor outer products. To do this, it uses the gate based model. Please, see Figure \ref{fig:QuantumCircuit} for an example of such a model. Polarisation of light can also be described using the mathematics introduced in the spin first approach \cite{Hennig2024Introducing}.

\begin{figure}[H]
\centering
\includegraphics[width=0.6\linewidth]{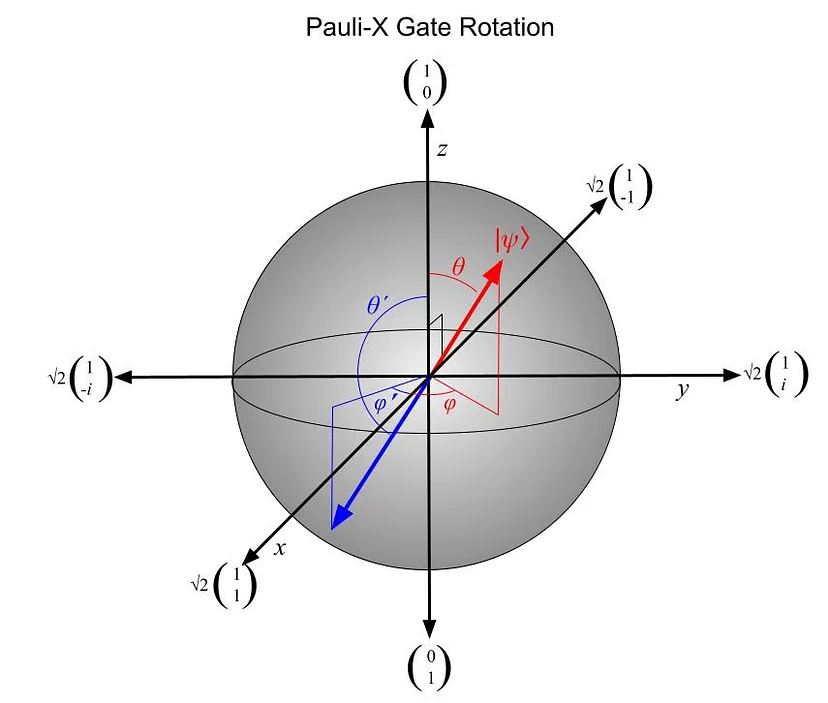}
\caption{\label{fig:SpinfirstBlochSphere} Image of a Bloch Sphere which is used to represent the qubit state vectors \cite{Shetty2025Visualising}. Specifically this diagram details the X Gate rotation which rotates a vector by 180 degrees around the X-axis. The specific vector expressions for states which lie on the axes are shown.}
\end{figure}

\begin{figure}[ht]
\centering
\includegraphics[width=0.6\linewidth]{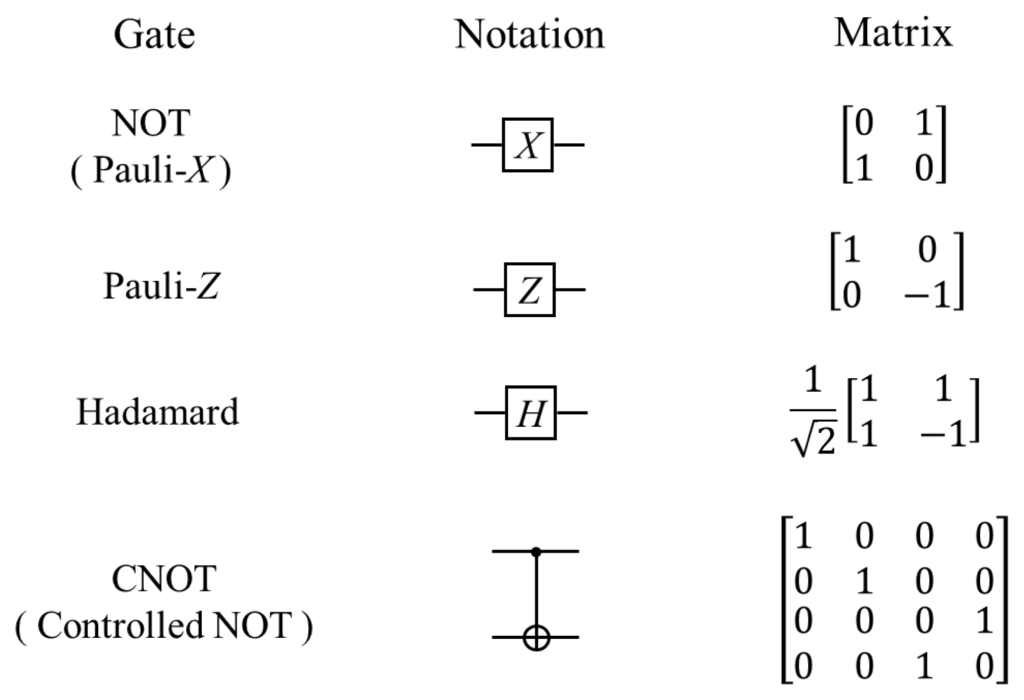}
\caption{\label{fig:SpinfirstGates} The gate matrices for several commonly used gates within a Quantum Circuit \cite{Yan2014Quantum}. In the Spin First Approach, these gates are applied to the vectors representing the different qubit states allowing for the vectors to be mathematically changed. These matrices are designed to perform a certain function on the qubits.} 
\end{figure}

\paragraph{} In the spin first approach, vectors are visualised on the Bloch Sphere and are used to represent qubits, see Figure \ref{fig:SpinfirstBlochSphere}. Gates then have the ability to reflect or rotate the vectors around the Bloch Sphere. Specifically, gates are introduced as unitary matrices, as shown in Figure \ref{fig:SpinfirstGates}. Single qubit gates are then represented as 2x2 matrices, and two qubit gates are represented as 4x4 matrices \cite{ERIC142}. Furthermore, entanglement, a core concept of quantum mechanics, is described as a qubit state not able to be represented with tensor products made up of single qubit states or on the Bloch Sphere. This enables most high school students to understand what occurs in the system when a quantum computer is running, without the need for the complex mathematics of wavefunctions.

\paragraph{} The American organisation QubitxQubit \cite{the_coding_school_qubitxqubit_2025} designed a program using the spin first approach.  
The QubitxQubit program aims to teach high school students and the general public through weekly lectures and computational labs, along with notebooks which include reference information, review, and coding exercises. The European Organization for Nuclear Physics (CERN) \cite{CERN} also curated and publicised a series of lectures using the spin first approach. These lectures had global reach and thus were able to educate students from many different backgrounds. Quantum coding packages are used within the hands-on coding portion of most of the spin first style courses, with the most common two being Quisket and Cirq.
Another notable spin first program is run by the Institute for Quantum Computing Canada which provides educational development for teachers as well as Spin first teacher resources for high school students \cite{Quantum101}.

\paragraph{} A strong advantage of the Spins First Approach is that it introduces students to key mathematical concepts relevant to quantum physics, such as bra-kets and Dirac Notation, at a level they can understand \cite{Merzel2024Mathematical}. Students are able to relate these mathematical concepts to key principles such as superposition and measurement \cite{Merzel2024Mathematical, Hennig2024Introducing}. Studying quantum computing courses which utilize the Spins First Approach allows students to realise how this emerging technology may be relevant to their future \cite{rasa_futurising_2022}. Furthermore, the Spins First Approach eliminates the need to understand complex physics topics, which allows students from varying backgrounds to participate \cite{arxiv001}. This supports a more diverse range of people to engage in Quantum Mechanics and Computing research and discovery.

\paragraph{} Despite the advantages and wide use of the Spins First Approach, it does possess disadvantages. The approach focuses on superposition of spins and therefore only allows for students to understand a limited number of quantum phenomena, excluding some key topics such as wavefunctions and tunneling. 
Furthermore, the Spins First Approach is more strongly centered on quantum information science rather than quantum physics. 
This can result in courses that utilise the Spins First Approach missing a range of other quantum phenomena important for understanding further topics such as manifestations. For example, this approach does not describe or mention discrete energy levels in atoms or the photo-electric effect, along with the discovery of photons. This creates gaps in student knowledge that do not allow them to fully understand the nuances of Quantum Computing and other Quantum Physics topics.

\section{Many Paths Approach to Quantum Mechanics}
\label{sec:many-parths}

The Many Paths Approach to Quantum Mechanics is based on Feynman's path integral formulation whereby quantum behavior is explained by assuming that particles simultaneously take all possible paths between two events.
 
Each path contributes a `probability arrow' that we call a phasor. The square of the sum of the phasors determines the overall likelihood of the final event. Despite some universities and education researchers pushing this approach since the 1970s, it has not been used widely. \\

\begin{figure}[ht]
\centering
\includegraphics[width=0.65\linewidth]{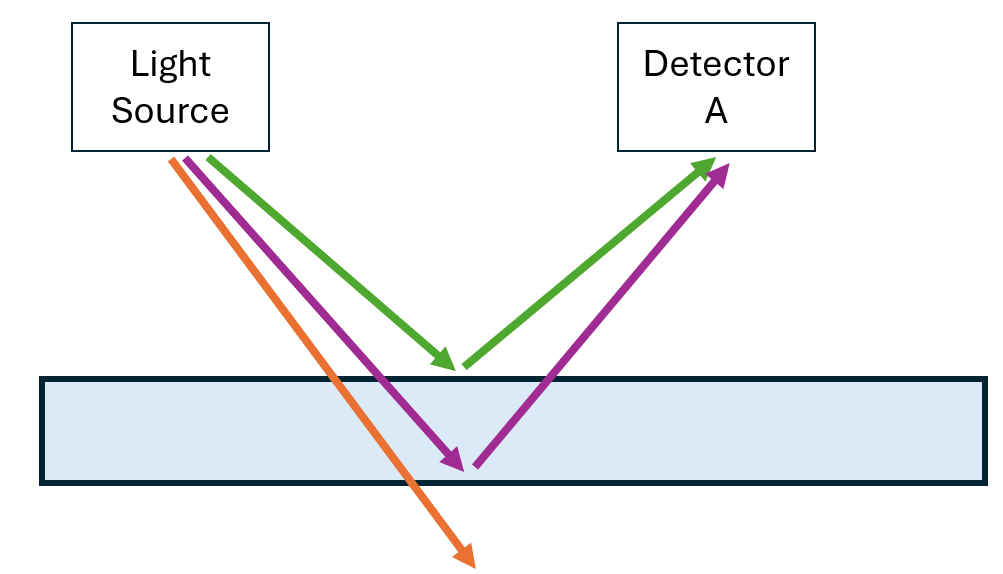}
\caption{\label{fig:QEDfig4} Diagram of thin film interference based on Figure 4 of QED \cite{FeynmanQED}. A light source emits a photon towards a thin piece of glass. For simplicity, we assume that the photon takes three paths, reflecting off the front surface of the glass (shown in green), reflecting off the back surface of the glass (shown in purple) or transmitting (shown in orange). The photon simultaneously takes all paths. 
The probability of a photon being detected at A varies with changing glass thickness as the paths illustrated in green and purple move between constructively or destructively interfering.}
\end{figure}

Richard Feynman introduces the Many Paths Approach at an accessible level in his book \textit{`QED The Strange Theory of Light and Matter'} \cite{FeynmanQED}. From this point on, we will refer to the book simply as QED. Feynman begins his explanation by asking the reader to imagine an experiment where a light source is shone onto a thin piece of glass. Each photon simultaneously follows the green, purple and orange paths shown in Figure \ref{fig:QEDfig4}. The phasors associated with each path are summed and squared to determine the total probability.\\

\begin{figure}[ht]
\centering
\includegraphics[width=0.65\linewidth]{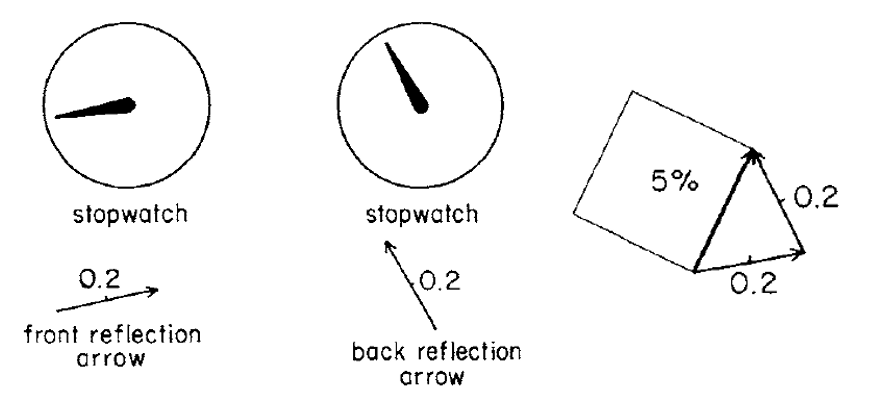}
\caption{\label{fig:QEDfig13} Diagram showing Feynman's visual model for calculating the probability of reflection of a photon \cite{FeynmanQED}. The phasor of the photon which reflects off the back of the glass rotates for a bit longer, due to the greater distance that it travels. The probability of reflection is found by adding the two phasors head to tail and finding the area of the corresponding square.}
\end{figure}

Feynman then uses phasors (the `Maths of Arrows') as part of a visual model to quantitatively explain this phenomenon. He considers two paths that a photon can take: one which reflects from the front surface and one which reflects off the back surface. 
We call the direction of the phasor, its \define{phase}. 
For each path, we start a stopwatch which represents the phase of the photon. As time passes, the phase rotates. 
We then add the phasors for both paths and square the result to find the probability of the photon reflecting, see Figure \ref{fig:QEDfig13}. 

Feynman then applies this same model to explain many other phenomena including plane mirror reflection, diffraction gratings, refraction, convex lenses, double slit interference and the effect of measurement. This approach can be extended to allow students to have one consistent conceptual model that explains quantum mechanics, classical mechanics (through the principle of stationary action and Euler-Lagrange equations \cite{McGinness2014Development}) and relativity (through the principle of maximal aging) \cite{Taylor2003Call}.\\

There are a number of resources that can assist instructors in teaching the Many Paths Approach. Richard Feynman introduces the approach in QED \cite{FeynmanQED}. A more extended treatment appropriate for undergraduate level can be found in the Feynman Lectures on Quantum Mechanics \cite{FeynmanLectures}. Many other instructors have also published teaching resources and methods for teaching these concepts to students
\cite{Taylor2008Principle,Gray2009Principle,Feynman2010Quantum,Taylor1998Teaching,Taylor2003Interactive,Swartz1972Physics,AAPT2012Least,McGinness2016Action,Malgieri2021Educational,Freericks2019Teaching} with the following being accessible at a high school level \cite{Ogborn2006Action,Cassiopela2009Principle,NCG1976Quantum,Fanaro2012Teaching} and this (Desmos) Geogebra simulation in Italian \cite{Malgieri2018Geogebra}.

In 2014, the Action Concept Inventory (ACI) was published to measure student understanding of the key concepts underlying the approach \cite{McGinness2014Development,McGinness2016Developing}. The ACI reveals that the four most common student misconceptions when taught the many paths approach are \cite{McGinness2014Development}:
\begin{enumerate}
  \item The number of paths available for an object to move from point A to point B increases as mass decreases.
  \item Small objects have more stationary paths than macroscopic objects.
  \item Each path has a probability with it and the probability is higher for the stationary path. 
  \item Macroscopic objects always take the stationary path, while quantum mechanical objects can `choose' different paths.
\end{enumerate}
Directly addressing misconceptions with students has been shown to be the most effective way of addressing them \cite{Randall2005Five}. However, further work needs to be done to confirm that this is effective for the Many Path Approach to Quantum Mechanics.\\ 

High School Teachers may find it difficult to introduce their students to the Many Paths Approach to Quantum Mechanics because it is not included in many curricula and there are few resources for teaching it at the high school level. The collection of resources referenced above should reduce this difficulty. The Many Paths Approach is conceptual and visual, which makes it an appealing approach for high school students. It can also be used to derive Newtonian mechanics, allowing students to visualise both quantum and classical mechanics in one consistent picture.

\section{Game-Based Quantum Learning}

Game-Based Quantum Learning refers to games, puzzles or simulations that students can complete which are designed to be both educational and entertaining. These can be used in the classroom, or students can engage with this individually in their own time. 
\cite{Douhauser2024Empirical} notices that many of the more “innovative social structures of activation strategies”, such as games, are exclusively aimed at high-school students to help them visualise the mechanics that students presented with. As many students tend to base their new learning of quantum mechanics upon classical mechanics due to prior learning, which often does not entirely quantify the behaviours that quantum particles exhibit. 

\paragraph{}The Game-Based learning approach is often paired with teacher resources and discussion prompts aimed at allowing the students to re-engage with the material after completing the game or subsequent levels. This group discussion approach to Gamified learning allows  students to have critical discussions with their peers and further engage with the material.
Game-Based Quantum Learning can generally be divided into three categories: online games, board or card games, and physical games. We define quantum games as games with one or more of the phenomena of quantum physics embedded in their game mechanics \cite{Proquest016}. This can include educational games aiming to directly introduce quantum mechanics to students or a general audience. Many quantum games are designed to teach students about a single quantum phenomenon, often using an analogy.

\paragraph{Online Games}
In 2024 Chiofalo et. al. \cite{Proquest016} found that there were more than 100 online quantum video/computer games which had been released, 41\% of which had a classical counterpart and 56\% which were developed in a hackathon. However, they classified only approximately $4\%$ of these games as educational, with most of the games being puzzles, simulations or arcade games. 
We direct the reader to \cite{Seskir2022Quantum} and \cite{Proquest016} for an overview of some of the most prominent games including `Quantum Moves 2', `The Alice Challenge' 
and `QPlayLearn'. \\

\begin{figure}[ht]
\centering
\includegraphics[width=0.8\linewidth]{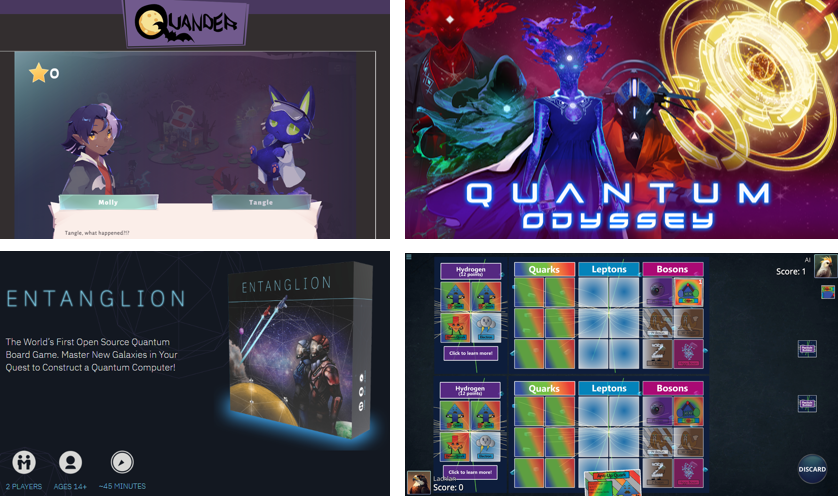}
\caption{\label{fig:QuantumGames} Images of different quantum games including `Quander', the `Quantum Odyssey' puzzle game, the board game `Entaglion' and the online version of `Particle Builder'.} 
\end{figure}

A team called Quarks Interactive have also developed a Game called Quantum Odyssey which allows users to learn about quantum circuits by solving a series of puzzles \cite{Eric020}. The authors performed a study (N=42) which found that students enjoyed the game, but did not get students to complete a pre/post-test to measure change in understanding \cite{Nita2021Inclusive}. Particle Builder \cite{Attar2025ParticleBuilder} is an example of an online game which aims to teach students multiple quantum mechanics concepts through particle physics including the Pauli Exclusion principle. A study (N=225) showed that students learned concepts directly related to the core game mechanics through playing the game \cite{McGinness2025ParticleBuilder}. Both Particle Builder and Quantum Odyssey are shown in Figure \ref{fig:QuantumGames}.

\paragraph{Board games and Card Games}
Many different physical quantum games have been developed for teaching quantum physics. One of the most popular is quantum tic-tac-toe with many variations of this game having been published \cite{Proquest010,Goff2022Quantum,Knight2021Pilot,Nagy2012Quantum,Weingartner2023Quantum}. Most studies into the effectiveness find (anecdotally) that playing the game improves student interest in quantum physics\cite{Proquest010}. 

Another example is Entanglion \cite{Entanglion}, a two-person cooperative board game which exposes students to the fundamental concepts of quantum computing such as superposition, entanglement, and gates \cite{IEEE004}. This has been used in conjunction with notebook style learning systems described in the Spins-First Section \ref{sec:SpinFirst} to teach quantum concepts to Year 9-12 students \cite{IEEE004}.

Escanez-Exposito et. al. developed a card game called `Qubit the Game' to foster enthusiasm for quantum computing and teach some foundational quantum information concepts \cite{ERIC040}. In the game students collect cards which represent gate operations and use these to put qubits into a desired sequence. By playing the game students learn about qubits, superposition, the Bloch Sphere, quantum gates, projection, entanglement and decoherence. The authors performed a study (n=96) and found that 16-17 year old students self-reported knowing more about quantum computing by playing the game \cite{ERIC040}.

\paragraph{Physical Games}

There are very few quantum games which are physical in nature. Entanglement ball is a variant of dodge-ball designed to teach students the concepts of quantum entanglement while enjoying physical exercise \cite{Eric156}. Anecdotal feedback indicated that students aged 13-18 years old enjoyed the game and developed some understanding of entanglement, but no thorough analysis has determined if students picked up misconceptions from playing the game \cite{Eric156}. There are also some physical objects such as Pauli cubes which students can play with to enhance their understanding of quantum information \cite{Gauvin2018Playing}.

In addition to physical games there are many studies which attempt to help students understand Quantum Circuits using 3D virtual reality simulations, see Maclean et. al. for an overview of these studies \cite{Maclean2022Virtual}. Overall there is a lack of conclusive evidence that there is significant educational benefit for students participating in these simulations \cite{Maclean2022Virtual}. 
Outside of VR, there are other visualisation tools such as QGrover which was designed to help students understand each of the steps in a quantum algorithm \cite{Norrie2024QGrover}. 

As well as quantum games, Chitransh et. al. introduced the general public, not just students, to quantum physics using theatre \cite{Chitransh2022Multidisciplinary}. The plays used analogies and metaphors to convey quantum concepts which could not be explained classically. Plays are also used at both the primary and secondary level as a part of the Einstein-First program \cite{Kaur2024Developing, Choudhary2018Can}, though little has been done to measure their effectiveness as a teaching tool \cite{Adams2021Long}.

\paragraph{Quantum Games in the Classroom}
The strengths of game-based quantum learning lie in the ability for students to experiment with the subject of quantum mechanics, as is often seen in school laboratories during classes. This ability to interact with the subject matter in  allows for a richer conceptual understanding than by theoretical and textbook learning on its own. Generally, mathematics is not needed for the Game based quantum learning approach, with concepts related to the topic being integrated into the game-play. \\

Although there are many different games which can cover many concepts, though they usually do not give students the same level of mathematical or historical depth as other methods \cite{Douhauser2024Empirical}. Games can be used to introduce students to particular concepts from quantum mechanics in an enjoyable and memorable way, but the games cannot serve as a complete unit by themselves. They can only be used to supplement other approaches to teaching quantum physics.

\section{Einstein-First Approach}

The Einstein-First Approach to teaching quantum mechanics was developed by the Einstein-First program \cite{EinsteinFirst} and is characterised by seven core principles: 1; hands-on activities, 2; use of toys, models and analogies, 3; activity based group learning, 4; whole-body learning, 5; appropriate language and keywords, 6; use of inexpensive equipment and 7; role-plays. 
The Einstein-First program aims to bring modern physics concepts, including quantum physics, general relativity, climate science, and astrophysics, to classrooms within existing physics curricula. 

\begin{figure}[ht]
\centering
\includegraphics[width=0.25\linewidth]{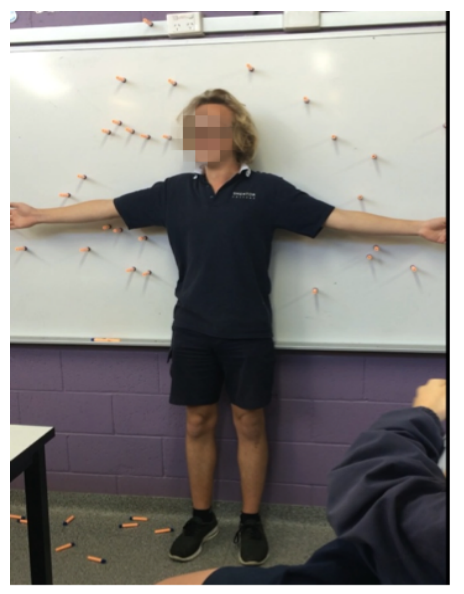}
\caption{\label{fig:BulletPhotography} A `silhouette photograph' of a student created by Nerf gun bullets, representing the linear trajectory of photons in flat space. It also demonstrates the particle-like properties of photons, namely their non-zero
momentum and energy. Figure reproduced from \cite{Kaur2017Teaching}.} 
\end{figure}

The Einstein-First program brings in quantum mechanical concepts relating to atoms and light in Year 3, Year 5 and Year 9. Figure \ref{fig:BulletPhotography} is an example activity which teaches students about both the wave and particle nature of light \cite{Kaur2017Teaching}. In this activity, Nerf bullets are used as analogies for photons, allowing students to visualise photography, light scattering, and the photoelectric effect. The wave-particle duality is illustrated by the model, referred to as its 'bulletiness' and 'waviness', respectively. The foundation laid by emphasising the particle nature of light is useful in further extension into concepts like photon momentum and uncertainty principle. 

The Einstein first approach makes use of active learning methods which result in an improved conceptual understanding of core principles \cite{Anttila2024Can}. The approach has been shown to have a positive impact on student attitudes towards science and physics \cite{Anttila2024Can}. The use models and analogies makes the approach visual and accessible to a range of students. Research into the outcomes of the Einstein-First program also suggests a strong benefit of the program in encouraging gender balance in STEM education and career choices \cite{Kaur2017Teaching}.

Despite positive impacts on student understanding and attitudes towards science, not all students will master the abstract concepts presented in Einstein First lessons \cite{Choudhary2018Can}, demonstrating the need for concepts to be revised and additionally, develop in complexity through their primary and secondary education. One of the greatest challenges in implementing the Einstein-First program is the requisite knowledge required by teachers so that they can identify and explain the problems with each analogy. This requires that teachers have access to sufficient training and resources, meaning that there must be significant teacher up-skilling.


\section{Conclusion}

We have provided a framework which allocates many approaches to teaching quantum mechanics one of seven categories: Historical Development, Defamiliarisation, Quantum Picturalism, Spin First, Many Paths, Game Based and, Einstein-First. Each of these areas focuses on teaching specific areas of quantum physics, using different methods.

An area for future work would be to expand this framework into a two-dimensional framework. The first dimension would be the content and concepts that are taught. For example many of the Game Based approaches attempt to teach many of the same concepts as the Quantum Picturalism and Spin First approaches. 
Although these three approaches cover similar ideas, they are very different in terms of the second dimension of the revised framework: the approaches used to teach the content. A two-dimensional framework would make it clearer for teachers the range of resources which are available to teach specific quantum content.

\section*{Acknowledgments}

We acknowledge the contribution of the Commonwealth Scientific and Industrial Research Organisation and the Australian National University for funding this research.

The ethical aspects of this research have been approved under ethics protocol 2023/1362 by the ANU Human Research Ethics Committee (HREC) of the Australian National University (ANU).

\bibliographystyle{splncs04}
\bibliography{CuriousMindsinQuantumReferences}

\end{document}